# When Familiarity Remains: Procedural Memory, Symbolic Anchors, and Digital Engagement in Dementia Care

*Jeongone Seo[1], Rutgers University, New Jersey, U.S.A*

*Kyung-zoon Hong, Sungkyunkwan University, Seoul, South Korea*

*Sol Baik, University of Virginia, Virginia, USA*

**INTRODUCTION:**

Older adults with early-stage dementia often retain procedural memory, enabling continued use of familiar technologies. Additionally, symbolic anchors—such as photos or personalized content—may serve as memory cues to reinforce digital engagement. This study explores how these mechanisms support technology use in dementia care within the South Korean context.

**METHODS:**

We conducted in-depth interviews with 11 professional caregivers of community-dwelling older adults with cognitive decline. Grounded theory methods guided the analysis, using iterative coding and constant comparison to identify emergent themes.

**RESULTS:**

Caregivers reported that familiar digital routines (e.g., taking photos) persisted through procedural memory. Symbolic anchors, such as family photos or recognizable icons, enhanced interaction and emotional engagement. However, unfamiliar or anthropomorphic technologies often triggered fear or symbolic resistance.

**DISCUSSION:**

Findings highlight the dual role of procedural memory and symbolic anchors in sustaining digital engagement. Designing culturally responsive and cognitively accessible technologies may enhance autonomy and well-being in dementia care.

# 1.Introduction

According to the National Health Insurance Service, over one million adults in South Korea received home-based care under the long-term care insurance system (LTCI)

---

in 2023 (1). This system serves both older adults and younger individuals with age-related conditions such as dementia. As the population ages rapidly, demand for care continues to grow, yet workforce shortages remain a chronic issue (2,3). In response, South Korea—widely recognized for its advanced technology infrastructure—has increasingly turned to ICT-based solutions, including smartphones, tablets, and AI-powered robots, to expand service delivery, manage care, and mitigate social isolation (4,5,6).

While these tools offer promise, integrating them into dementia care presents complexities. Cognitive impairments often limit digital use due to memory loss, motor decline, and low device accessibility (7,8,9). Technology anxiety and low digital literacy further compound these barriers (10,11). Yet, non-use is not always rooted in inability; it can reflect intentional disengagement from content perceived as overwhelming or irrelevant (12). Understanding how technology can be more meaningfully embedded in the lives of people with dementia requires attention to both cognitive and symbolic mechanisms.

Procedural memory—responsible for learned routines—often remains relatively intact in early dementia and may support continued interaction with familiar technologies (13,14). Older adults may still perform tasks like taking photos or checking the weather, even if semantic comprehension is impaired. This raises questions about how prior experience and habit formation affect digital persistence.

Beyond memory processes, symbolic anchors—such as photos, music, or familiar icons—can evoke recognition and emotional engagement. These anchors help sustain digital habits and promote interaction, particularly when embedded into personalized or culturally resonant content (15,16). ICT-based interventions often integrate such elements as memory cues to support user interaction and caregiver connection (17,18). These cues serve both emotional and functional roles, enhancing usability and promoting sustained engagement.

Prior studies suggest that digital tools embedded in daily routines—like photo-sharing interfaces or touch-based devices—can become "habitual artifacts" that reinforce identity and continuity in dementia care (19). Similarly, emotionally resonant technologies, such as social robots and virtual environments, have demonstrated symbolic and mood-enhancing effects (20). However, most of this work centers on Western contexts and rarely addresses how procedural memory and symbolic anchoring operate jointly.

There is a growing need to explore these mechanisms in non-Western settings. In Korea, where older adults show high smartphone penetration, digital engagement is often shaped by collectivist values, family dynamics, and community interdependence (12). Understanding how digital habits are maintained—or lost—after cognitive decline in such cultural settings can inform more inclusive and responsive design.

While concepts like interpretive gatekeeping have been used to study intergenerational technology use (21), few studies apply them to dementia care. Recent research urges designers to accommodate cognitive diversity by embedding personal, social, and cultural contexts into technology systems for people with dementia (22,23,24).

To address these gaps, this study asks: (1) How do older adults with cognitive decline maintain or lose digital habits, and what role does procedural memory play? (2) What impact do symbolic anchors—like personalized content and self-referential cues—have on sustaining digital engagement? Findings aim to inform culturally sensitive, cognitively accessible technologies that promote autonomy, connection, and well-being in dementia care.

## 2. Methods

### 2.1 Study Design

This qualitative study employed a constructivist grounded theory approach (25) to examine how Korean care professionals interpret and support the use of digital technology among people experiencing early-stage dementia. Grounded theory was chosen for its ability to generate theoretical insights from lived experiences and symbolic interactions, allowing the researcher to explore the interplay of procedural memory, symbolic anchors, anxiety, and caregiving support in sustaining digital engagement.

### 2.2 Setting and Participants

Participants were recruited from community-based care environments in South Korea, where digital tools are increasingly integrated into services for adults with cognitive decline. All participants were either current or former staff at government-certified adult day care centers, a key service model within Korea's public long-term care insurance (LTCI) system. These centers provide structured, daytime support for older adults who have been assigned a care level based on standardized cognitive and functional assessments under the LTCI eligibility criteria (26). As outlined in recent analyses, adult day care centers are essential components of Korea's effort to promote aging-in-place, reduce caregiver burden, and prevent premature institutionalization, particularly for individuals with dementia (27). Moreover, validated assessment tools developed specifically for these settings underscore the expertise of care workers in interpreting behavioral and functional changes among older adults with cognitive impairment (28). As such, participants in this study were well-positioned to speak to the digital experiences, anxieties, and adaptive behaviors of cognitively impaired older adults in real-world service delivery contexts.

After receiving ethical approval from the Institutional Review Board of Sungkyunkwan University (IRB No. 2023-01-014), the study employed purposive and snowball sampling to recruit 11 care professionals, including home care aides, support staff, social workers, and a program director. Ten participants identified as women and one as a man, with ages ranging from their early 40s to late 60s and caregiving experience spanning from 5 months to 7 years.

Table 1. Overview of Caregiver Participants in the Dementia Study

| Participant | Gender | Age | Experience | Role |
| --- | --- | --- | --- | --- |
| Participant1 | M | Late 60s | 2 years | Care worker (Transportation support) |
| Participant2 | F | Mid 50s | 6 months | Social worker |
| Participant3 | F | Early 50s | 5 years | Care worker |
| Participant4 | F | Late 40s | 5 years | Daily support worker/Care worker |
| Participant5 | F | Early 50s | 1 year | Assistant nurse |
| Participant6 | F | 50s | 5 years | Care worker |
| Participant7 | F | Early 50s | 5 months | Social worker |
| Participant8 | F | Early 50s | 1+ year | Assistant nurse |
| Participant9 | F | Late 40s | 2 years | Daily support worker |
| Participant10 | F | 50s | 4 years | Care worker |
| Participant11 | F | 50s | 7 years | Center director, social worker |

## 2.3 Data Collection

Semi-structured interviews were conducted by phone in Korean between September and October 2024. Each session lasted between 40 and 60 minutes and was audio-recorded with consent. The interview protocol explored participants' caregiving routines, interpretive strategies, emotional responses, and insights regarding how individuals with dementia engage with digital technologies. Participants received an honorarium of KRW 30,000 for their participation.

## 2.4 Data Analysis

Interview transcripts were analyzed using the iterative coding procedures of constructivist grounded theory (29,30,31):

- Initial coding: Line-by-line coding of transcripts to identify actions, meanings, and experiences.

- Focused coding: Consolidation of key codes into thematic categories.

- Axial coding: Analysis of relationships among categories to uncover contextual conditions and social processes.

- Theoretical integration: Development of a conceptual framework grounded in the data and informed by cultural context.

While the analysis began with sensitizing concepts such as procedural memory and symbolic cues, the process remained open to unanticipated insights, such as emotional resistance to unfamiliar AI tools or the role of family members as digital intermediaries. Constant comparative analysis and memo-writing were used to refine emerging categories and ensure analytic rigor (32).

## 2.5 Trustworthiness and Reflexivity

Several strategies were used to enhance the credibility and transparency of the findings (33):

- Member checks: Preliminary analytic summaries were shared with participants for validation.

- Peer debriefing: The researcher engaged in regular discussions with the academic advisor to support analytic clarity.

- Thick description: Contextual details were preserved to enhance transferability.

- Audit trail: A complete record of coding steps, memos, and analytical decisions was maintained.

The researcher brought a dual perspective as both a social work practitioner and doctoral researcher. This positionality enabled interpretive depth while also requiring ongoing reflexive attention to minimize bias.

## 2.6 Ethical Considerations

The study adhered to the principles outlined in the Declaration of Helsinki. All participants were informed of the purpose of the study, their right to withdraw at any time, and the measures taken to ensure confidentiality. Verbal and written informed consent was obtained. All identifying information was removed from transcripts, and data were stored securely in encrypted files accessible only to the research team.

## 3. Results

This study explores how older adults with cognitive decline engage with digital technologies within a cultural and social ecology shaped by memory, caregiving, and meaning-making. The core category that emerged from the data is the tension between sustaining digital routines through familiarity versus disengagement due to strangeness. This phenomenon is illuminated through five interrelated components: causal conditions, contextual conditions, intervening conditions, actions and strategies, and consequences.

Table 2. Conceptual Categories and Subcategories Explaining Digital Engagement Among Older Adults With Cognitive Decline

| Category | Subcategory | Conceptual Meaning |
|---|---|---|
| Causal Conditions | Embodied familiarity | Long-term digital habits support procedural memory use |
| | Barriers to late adoption | Lack of early experience makes late adoption hard |
| | Motivational disposition | Desire for continuity affects willingness to learn |
| | Technological trauma | Negative experiences lead to digital withdrawal |
| Contextual Conditions | Symbolic fears | Cultural beliefs shape fear of anthropomorphic devices |
| | Surveillance concerns | Privacy fears result in rejection of tech devices |
| | Trusted social spaces | Learning is enabled in emotionally safe social contexts |
| Actions & Strategies | Routine reinforcement | Repetition of familiar actions sustains digital use |
| | Symbolic anchors | Personal content (names, photos) enhances engagement |
| | Avoidance behaviors | Rejection behaviors to regain control over discomfort |
| | Caregivers as proxies | Caregivers act as emotional and technical mediators |
| Intervening Conditions | Device simplicity | Simple, consistent UI helps maintain digital routines |
| | Caregiver temperament | Patience and empathy in caregivers reinforce learning |
| | Peer-based learning | Group learning reduces shame and supports adaptation |
| Consequences | Preserved autonomy | Digital use maintains identity and connection |
| | Delegated use | Caregivers gradually take over digital roles |
| | Defensive withdrawal | Complete rejection due to fear and mistrust |

Together, they map the complex sociocognitive terrain in which digital interaction unfolds among individuals experiencing cognitive decline.

### 3.1 Causal Conditions

*Embodied Familiarity and Procedural Memory*

A central factor shaping digital engagement among older adults with cognitive decline was prior digital familiarity. Participants consistently emphasized that repeated, long-term use of specific digital tools before the onset of dementia often allowed continued interaction through procedural memory. Unlike declarative memory, which tends to deteriorate early in dementia, procedural memory supports embodied routines that persist longer.

Participant 5 explained, "If it's something that's been used since before the onset of dementia, I don't think he'll forget it." Participant 9 described an 87-year-old client who, despite significant cognitive decline, could still check train schedules and send messages to family on KakaoTalk: "It's what he's always done. His fingers just know what to do." These accounts highlight how ingrained behavioral patterns—formed over years—enable older adults to navigate digital environments even as other cognitive functions erode.

*Barriers of Late-Life Digital Adoption*

Conversely, the absence of prior digital experience posed profound barriers to engagement. Several participants noted that older adults who had not used smartphones or digital interfaces earlier in life found it nearly impossible to begin doing so after cognitive decline set in. Participant 4 explained, "If they never used a phone before and are suddenly told to make a call, it is often not possible." Others echoed this point, noting that even simple tasks like adjusting volume or navigating a home screen could become insurmountable in the absence of foundational familiarity. This suggests that cognitive decline alone does not explain disengagement—rather, it is compounded by the lack of prior embodied familiarity.

*Motivational Disposition Toward Technology*

Motivation also emerged as a critical causal factor. Some older adults expressed curiosity about new features, while others showed emotional fatigue or ambivalence. The desire to "keep living the way they've always lived" often overrode the effort required to engage with new tools. Participant 5 reflected, "Most would rather just continue living the way they've always lived." Participant 6 added that "when they feel like it's something young people do, they don't even try." These comments reflect a desire for stability and continuity, particularly as cognitive and emotional resources become more fragile with age.

*Technological Trauma and Emotional Withdrawal*

Past negative experiences further inhibited engagement. Several participants described instances in which older adults withdrew from technology use after episodes of confusion or scams. In such cases, digital withdrawal appeared to serve as a protective strategy, allowing individuals to avoid perceived threats and reassert control by delegating digital tasks to trusted caregivers. Participant 1 recounted how her mother, after falling victim to a phishing scam, completely withdrew from using her phone. "She said, 'You do it.' She doesn't even want to touch it now."

### 3.2 Contextual Conditions

*Symbolic Fears Toward Anthropomorphic Devices*

Individual resistance was deeply shaped by a broader cultural and technological environment. Participants often described older adults' emotional responses to unfamiliar devices as shaped by symbolic meanings and generational beliefs, rather than by functionality alone. Participant 9 described a case where a client covered an AI doll with a blanket, saying, "They said the AI doll was really scary. They were afraid a ghost might enter it." These responses, while seemingly irrational, reflected deeply rooted spiritual beliefs and discomfort with anthropomorphized technology.

*Perceived Surveillance and Loss of Privacy*

Participants also shared numerous accounts of older adults rejecting devices that felt invasive. Participant 3 recalled a woman who said, "It's staring at me and watching everything," referring to a motion sensor installed in her room. She eventually covered the device and refused to sleep near it. Participant 1 described another case in which clients refused ward-distributed smart speakers because "they thought it was like CCTV. They didn't want to be watched." Such narratives show how fears of surveillance and the erosion of privacy can undermine well-intentioned interventions.

*The Role of Trusted Social Spaces for Learning*

Despite these anxieties, participants also identified contexts where older adults were more receptive to digital learning. Socially embedded environments such as senior centers, churches, and cultural centers provided emotional safety and community reinforcement. Participant 2 noted that her mother only began using search functions on her phone "after she learned it at a cultural center with her friends. She felt safe there." These spaces offered more than skill instruction—they provided affective buffers and collective encouragement, reducing anxiety and reframing technology as a shared experience.

### 3.3 Actions and Strategies

*Routine Reinforcement Through Familiar Behaviors*

In response to both internal dispositions and contextual challenges, older adults and caregivers employed a range of strategies to sustain digital interaction. One common method was reinforcing previously established routines. Participant 1 shared, "My mother still listens to music and looks up church videos on YouTube. She doesn't think—it's just automatic." Similarly, Participant 2 described how church members "send pictures to their children on KakaoTalk because they've been doing it for years." These routines were less about active learning and more about sustaining continuity through procedural memory.

*Symbolic Anchors and Self-Referential Design*

Another frequently used strategy was leveraging symbolic content to anchor attention and recognition. Participant 7 emphasized, "They pay attention only if their name or their kids' names come up. If it's not personal, they ignore it." Participant 2 noted that her mother "felt proud sending family pictures. It wasn't just using the phone—it meant something." These emotionally resonant elements—names, faces, music—served as bridges between digital interaction and the user's personal identity, making the technology feel familiar and worthwhile.

*Avoidance Behaviors in the Face of Discomfort*

However, when discomfort overwhelmed recognition, avoidance became the dominant response. Participant 3 recalled, "One client covered the smart speaker with a scarf. She didn't want it looking at her." In other cases, users would move devices into drawers or unplug them entirely. These were not passive refusals but active efforts to reassert control over perceived intrusions. Participant 1 observed, "Some just hand you the phone and say, 'You do it.' They've made the decision to opt out."

*Caregivers as Mediators and Proxies*

In such cases, caregivers became critical digital intermediaries. Participant 1 explained, "You have to teach it one by one, again and again." Her role often extended beyond instruction to emotional reassurance. "Even if you teach it again... you have to do it again. You have to be patient. They feel ashamed if they forget." These caregiving roles involved not just technical help, but affective labor—encouraging, calming, and validating. Over time, many caregivers assumed partial or full responsibility for digital tasks, enabling continuity even when direct use was relinquished.

### 3.4 Intervening Conditions

*Device Simplicity and Consistency*

The success of these strategies was mediated by several intervening conditions that either enabled or constrained engagement. Simplicity of device interface emerged as a key factor. Participant 5 emphasized that "the functions should be simple, and if it is used repeatedly, it becomes easier to remember." Overly complex or dynamic systems, particularly those with frequent software updates, disrupted learned routines and created disorientation. Participants noted that once familiar icons or button placements changed, older adults often felt lost and gave up.

*Emotional Temperament of Caregivers*

Equally crucial was the caregiver's emotional disposition. Caregivers who exhibited patience, consistency, and a calm presence fostered a sense of psychological safety.

Participant 1 underscored this, stating, "Even if you teach it again... you have to do it again." The capacity to offer non-judgmental, repeated support was often the difference between limited use and complete rejection. Emotional scaffolding thus played a fundamental role in enabling fragile forms of digital connection.

*Shared Learning as a Social Reinforcer*

The learning environment itself also mattered. Participant 6 observed that older adults adapted more effectively when learning alongside peers: "They learn better when they learn with people of similar ages. It becomes a shared experience, not a solitary struggle." Group settings allowed for normalization of mistakes, mutual encouragement, and intergenerational bridging. These intervening conditions—simplicity, emotional support, and communal learning—helped translate abstract strategies into concrete outcomes.

### 3.5 Consequences

*Preservation of Autonomy Through Digital Familiarity*

The interplay of causal, contextual, and intervening factors produced a range of outcomes, from sustained autonomy to total withdrawal. In the most positive cases, older adults retained digital agency despite cognitive decline. Participant 5 described a man who could still independently call his son using his mobile phone, while Participant 2 emphasized that her mother took pride in sending music or images to family members. These actions, though modest, served as affirmations of continuity and control.

*Long-Term Delegation and Agency Loss*

However, not all outcomes were empowering. When older adults delegated digital responsibilities, caregivers became long-term proxies, resulting in a transfer of agency. What began as a short-term coping strategy became habitual dependency. Participant 1's case exemplifies this: "She just throws the phone to me." This dynamic can offer short-term efficiency but potentially erodes long-term autonomy, especially when it becomes routinized.

*Digital Disconnection as Defensive Withdrawal*

In more extreme cases, engagement was completely severed. The fear of being monitored led some to reject digital tools outright. Participant 1 recalled clients who refused to install ward-distributed smart sensors, citing surveillance concerns. This avoidance resulted in disconnection not only from digital services but also from opportunities for social contact and health support. The consequence, ultimately, was isolation reinforced by technology.

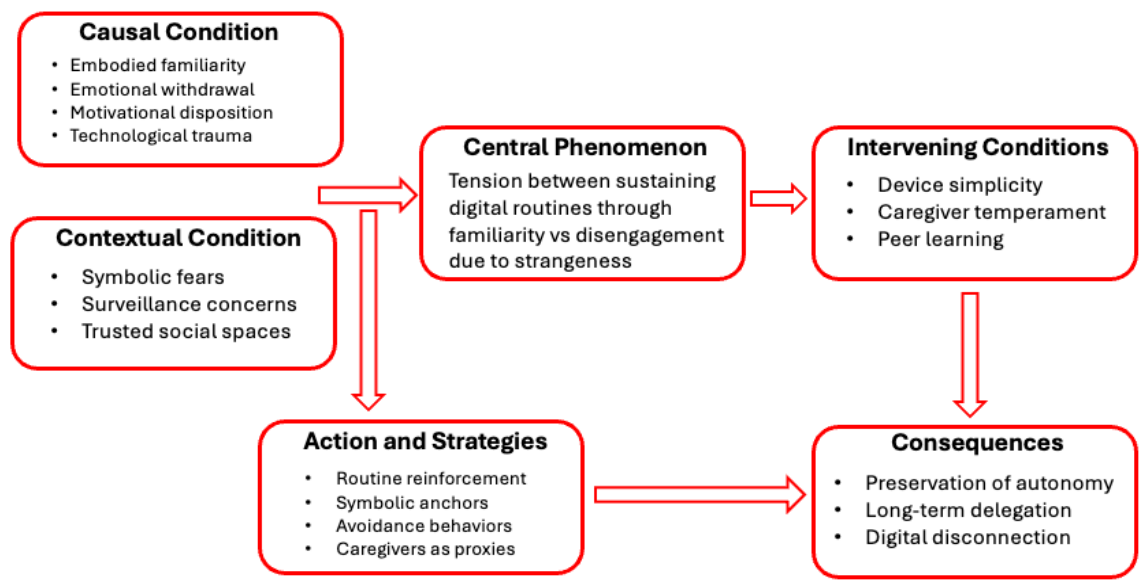

Figure 1. Conceptual Paradigm Model of Digital Engagement in Cognitive Decline

Figure 1 provides a visual synthesis of the five interconnected components, illustrating how causal, contextual, and intervening conditions interact with strategies and consequences in shaping digital engagement among older adults with cognitive decline.

## 4. Discussion

This study provides new insights into the layered processes by which older adults with cognitive decline maintain, alter, or disengage from digital engagement, especially in non-Western contexts such as South Korea. Drawing on the core concepts of procedural memory and symbolic anchoring, we have shown that digital routines can persist when cognitive, emotional, and social supports converge. The findings challenge dominant narratives of older adults as passive or incapable digital users by foregrounding the dynamic interplay of memory, motivation, cultural beliefs, and caregiving support.

**4.1 Interpretation of Key Findings**

4.1.1 The Dual Role of Memory: Procedural Routines and Symbolic Anchors

Our findings affirm the critical role of procedural memory in sustaining digital interaction, even as other cognitive functions decline. Repeated exposure and habitual use were shown to preserve embodied routines, allowing older adults to continue using familiar applications such as KakaoTalk or music apps. This aligns with prior research that highlights the relative resilience of procedural memory in early dementia (13,14). Importantly, our participants did not merely report continued use, but framed it as a form of agency—enabling them to meet personal or social needs despite cognitive barriers.

The second key mechanism sustaining digital engagement was symbolic anchoring, often mediated through culturally meaningful content. Personal photographs, familiar names, and emotionally resonant music served not only as memory cues but also as sources of pride, recognition, and connection. These findings extend previous work (17,18) by showing that symbolic elements do more than facilitate interaction—they serve as relational infrastructure. In the South Korean context, where family ties and spiritual interpretations shape emotional experience, symbolic anchoring bridges the gap between usability and meaning.

These findings resonate with Lazar and Kirk's concept of *habitual artefacts*, emphasizing how technologies used in consistent, emotionally salient ways can stabilize identity and reinforce routine engagement even in the face of memory loss (19).

4.1.2 Beyond the Individual: The Caregiver as an Interpretive Mediator

A notable empirical contribution of this study is the documentation of caregivers' emotional labor and interpretive work. Their role extends beyond technical assistance; they act as translators, encouragers, and sometimes digital surrogates. This aligns with the concept of interpretive gatekeeping, in which a more technologically or socially empowered family member mediates technology use not only functionally but also symbolically—filtering content through cultural, emotional, or ethical frameworks (21).

Furthermore, we found that digital uptake was enabled by relational and institutional support—especially through community centers and caregiver mediation. This suggests a model of socio-technical learning that prioritizes interdependence and safe spaces. Group learning among peers created a social buffer against the shame or anxiety often associated with technology adoption, functioning not only as sources of knowledge but as affective scaffolds.

4.1.3 A Cultural Reinterpretation of "Non-Use"

The study affirms the need for culturally situated design in dementia-friendly technologies. Korean older adults' fear of surveillance and anthropomorphic devices reflected a symbolic logic shaped by folk beliefs and societal discourse. Technologies perceived as "watching" or "possessed" were often rejected outright. These responses highlight that merely simplifying interfaces may be insufficient when devices conflict with deeper cultural scripts. This heterogeneity calls for a move beyond binary classifications of "users" and "non-users" and instead supports a more nuanced continuum-based view of digital participation in dementia contexts.

**4.2 Implications**

4.2.1 Implications for Design: Memory-Sensitive and Emotion-Centered Principles

The results call for design principles that go beyond accessibility to include emotional resonance and memory compatibility. Interfaces should support routine-based interaction, allow for personalization, and embed symbolic content—such as emotionally meaningful media—that reflects the user's identity (15,16).

Additionally, devices should be able to accommodate fluctuation in engagement—recognizing that a person's willingness and ability to interact may vary depending on emotional state, contextual triggers, and cognitive condition.

Moyle et al. further illustrate how emotionally evocative design—such as virtual forests or pet-like companions—can trigger symbolic recognition and promote psychological well-being among people with dementia (20). Together, these findings highlight the need to view technology not just as a tool for task completion, but as an emotionally and culturally mediated artifact.

4.2.2 Implications for Care Practice and Policy

Our findings confirm that older adults often rely on interdependent strategies to manage digital life, such as delegation to trusted caregivers. While this proxy use can preserve short-term functionality, it may also erode long-term autonomy when it becomes routinized. This points to the need for design and care frameworks that accommodate varying levels of participation, including shared use models that recognize interdependence without defaulting to dependency. Supporting the emotional labor of caregivers and promoting group learning in trusted community spaces are also key policy considerations.

**4.3 Limitations and Future Directions**

Several limitations must be acknowledged. First, the sample was limited to professional caregivers in South Korea and did not include the voices of older adults themselves. This indirect approach may limit the depth of first-person narratives. However, all participants were current or former staff at government-certified adult day care centers, a core service under Korea's public long-term care insurance system. These centers exclusively serve older adults who have been assigned an official care level based on cognitive and functional assessments, including those diagnosed with dementia. As such, participants had sustained, firsthand exposure to the behavioral, emotional, and technological responses of older adults with cognitive impairment in structured care settings. Their interpretations, while indirect, reflect accumulated expertise in observing real-world engagement patterns. Future research should incorporate direct accounts from people living with dementia to triangulate caregiver perspectives.

Second, this study focused primarily on early-stage cognitive decline; more work is needed to understand digital interaction in moderate to late-stage dementia. Third, while the findings are culturally specific, they raise transferable questions for design in other high-tech but aging societies.

Future research could expand on this work by (1) conducting comparative studies across cultural contexts; (2) co-designing digital tools with caregivers and users living with dementia; and (3) exploring long-term outcomes of symbolic design interventions on quality of life and emotional wellbeing.

## 5. Conclusion

By centering memory, meaning, and mediated relationships, we can reimagine digital inclusion not as a binary goal, but as a dynamic and contextually grounded process. This study contributes a nuanced, culturally-situated model for understanding technology engagement among people with cognitive decline, highlighting that technology is not merely a tool, but a critical site where autonomy, identity, and connection are negotiated.

**Acknowledgments**

Jeongone Seo received research funding from the Korea Health Industry Development Institute (KHIDI) [grant number HI22C1477] for the period 2023–2025. The other authors declare no relationships, activities, or interests related to the content of this manuscript. Part of the data analyzed in this article was originally collected as part of the author's doctoral dissertation at Sungkyunkwan University, approved by the university's IRB (IRB No. 2023-01-014).

Portions of this manuscript (e.g., language refinement, paraphrasing) were supported using generative AI tools (ChatGPT by OpenAI), under the direction of the first author. The first author assumes sole responsibility for the use of these tools, including any potential errors or inaccuracies resulting from their use.

**Conflict of interest statement**

The author declares no conflicts of interest related to this study.

**Consent statement**

All participants provided informed consent prior to participation. The study was approved by the Institutional Review Board of Sungkyunkwan University (IRB No. 2023-01-014), and all procedures complied with the ethical standards outlined in the Declaration of Helsinki.